\documentclass[aip,rsi,superscriptaddress,reprint,floatfix]{revtex4-2}

\usepackage{graphicx}
\usepackage{dcolumn}
\usepackage{bm}
\usepackage[utf8]{inputenc}
\usepackage[T1]{fontenc}
\usepackage{mathptmx}
\usepackage{etoolbox}
\usepackage{xcolor}
\usepackage{amsmath}
\usepackage{siunitx}
\usepackage{textcomp}
\usepackage[colorlinks=true,allcolors=blue]{hyperref}
\usepackage{lineno}

\usepackage{tikz}
\definecolor{lime}{HTML}{A6CE39}
\DeclareRobustCommand{\orcidicon}{
	\begin{tikzpicture}
	\draw[lime, fill=lime] (0,0) 
	circle [radius=0.16] 
	node[white] {{\fontfamily{qag}\selectfont \tiny ID}};
	\draw[white, fill=white] (-0.0625,0.095) 
	circle [radius=0.007];
	\end{tikzpicture}
	\hspace{-2mm}
}
\foreach \x in {A, ..., Z}{\expandafter\xdef\csname orcid\x\endcsname{\noexpand\href{https://orcid.org/\csname orcidauthor\x\endcsname}
			{\noexpand\orcidicon}}
}

\begin{document}

\title{Achieving ultra-low and -uniform residual magnetic fields in a very large magnetically shielded room for fundamental physics experiments}

\author{N.\,J.~Ayres}
\affiliation{
Institute for Particle Physics and Astrophysics,
ETH Z\"urich,
CH-8093 Zurich, Switzerland}

\author{G.~Ban}
\affiliation{Normandie Universit\'e, ENSICAEN, UNICAEN, CNRS/IN2P3, LPC Caen, 14000 Caen, France}

\author{G.~Bison\orcidK{}}
\thanks{\bf The author to whom correspondence may be addressed: efrain.segarra@psi.ch, georg.bison@psi.ch}
\affiliation{Paul Scherrer Institut, CH-5232 Villigen PSI, Switzerland}

\author{K.~Bodek}
\affiliation{Marian Smoluchowski Institute of Physics, Jagiellonian University, 30-348 Cracow, Poland}

\author{V.~Bondar}
\affiliation{
Institute for Particle Physics and Astrophysics,
ETH Z\"urich,
CH-8093 Zurich, Switzerland}

\author{T.~Bouillaud}
\affiliation{Universit\'e Grenoble Alpes, CNRS, Grenoble INP, LPSC-IN2P3, 38026 Grenoble, France}

\author{D.~Bowles}
\affiliation{Department of Physics and Astronomy, University of Kentucky, Lexington 40506, Kentucky, USA}

\author{E.~Chanel}
\affiliation{Laboratory for High Energy Physics and
Albert Einstein Center for Fundamental Physics,
University of Bern, CH-3012 Bern, Switzerland}

\author{W.~Chen}
\affiliation{
Institute for Particle Physics and Astrophysics,
ETH Z\"urich,
CH-8093 Zurich, Switzerland}
\affiliation{Paul Scherrer Institut, CH-5232 Villigen PSI, Switzerland}

\author{P.-J.~Chiu\orcidE{}}
\affiliation{
University of Z\"urich,
CH-8057 Zurich, Switzerland}

\author{C.\,B.~Crawford}
\affiliation{Department of Physics and Astronomy, University of Kentucky, Lexington 40506, Kentucky, USA}

\author{O.~Naviliat-Cuncic}
\affiliation{Normandie Universit\'e, ENSICAEN, UNICAEN, CNRS/IN2P3, LPC Caen, 14000 Caen, France}

\author{C.\,B.~Doorenbos}
\affiliation{
Institute for Particle Physics and Astrophysics,
ETH Z\"urich,
CH-8093 Zurich, Switzerland}
\affiliation{Paul Scherrer Institut, CH-5232 Villigen PSI, Switzerland}

\author{S.~Emmenegger}
\affiliation{
Institute for Particle Physics and Astrophysics,
ETH Z\"urich,
CH-8093 Zurich, Switzerland}

\author{M.~Fertl}
\affiliation{Institute of Physics, Johannes Gutenberg University, D-55128 Mainz, Germany}

\author{A.~Fratangelo}
\affiliation{Laboratory for High Energy Physics and
Albert Einstein Center for Fundamental Physics,
University of Bern, CH-3012 Bern, Switzerland}

\author{W.\,C.~Griffith}
\affiliation{Department of Physics and Astronomy, University of Sussex, Falmer, Brighton BN1 9QH, United Kingdom}

\author{Z.\,D.~Grujic}
\affiliation{Institute of Physics, Photonics Center, University of Belgrade, 11080 Belgrade, Serbia}

\author{P.\,G.~Harris}
\affiliation{Department of Physics and Astronomy, University of Sussex, Falmer, Brighton BN1 9QH, United Kingdom}

\author{K.~Kirch\orcidB{}}
\affiliation{
Institute for Particle Physics and Astrophysics,
ETH Z\"urich,
CH-8093 Zurich, Switzerland}
\affiliation{Paul Scherrer Institut, CH-5232 Villigen PSI, Switzerland}

\author{V.~Kletzl\orcidF{}}
\affiliation{
Institute for Particle Physics and Astrophysics,
ETH Z\"urich,
CH-8093 Zurich, Switzerland}
\affiliation{Paul Scherrer Institut, CH-5232 Villigen PSI, Switzerland}

\author{J.~Krempel}
\affiliation{
Institute for Particle Physics and Astrophysics,
ETH Z\"urich,
CH-8093 Zurich, Switzerland}

\author{B.~Lauss\orcidG{}}
\affiliation{Paul Scherrer Institut, CH-5232 Villigen PSI, Switzerland}

\author{T.~Lefort}
\affiliation{Normandie Universit\'e, ENSICAEN, UNICAEN, CNRS/IN2P3, LPC Caen, 14000 Caen, France}

\author{A.~Lejuez}
\affiliation{Normandie Universit\'e, ENSICAEN, UNICAEN, CNRS/IN2P3, LPC Caen, 14000 Caen, France}

\author{R.~Li}
\affiliation{Instituut voor Kern- en Stralingsfysica, University of Leuven, B-3001 Leuven, Belgium}

\author{P.~Mullan\orcidL{}}
\affiliation{
Institute for Particle Physics and Astrophysics,
ETH Z\"urich,
CH-8093 Zurich, Switzerland}

\author{S.~Pacura}
\affiliation{Marian Smoluchowski Institute of Physics, Jagiellonian University, 30-348 Cracow, Poland}

\author{D.~Pais}
\affiliation{
Institute for Particle Physics and Astrophysics,
ETH Z\"urich,
CH-8093 Zurich, Switzerland}
\affiliation{Paul Scherrer Institut, CH-5232 Villigen PSI, Switzerland}

\author{F.\,M.~Piegsa}
\affiliation{Laboratory for High Energy Physics and
Albert Einstein Center for Fundamental Physics,
University of Bern, CH-3012 Bern, Switzerland}

\author{I.~Rien\"acker}
\affiliation{Paul Scherrer Institut, CH-5232 Villigen PSI, Switzerland}

\author{D.~Ries}
\affiliation{Paul Scherrer Institut, CH-5232 Villigen PSI, Switzerland}

\author{G.~Pignol\orcidJ{}}
\affiliation{Universit\'e Grenoble Alpes, CNRS, Grenoble INP, LPSC-IN2P3, 38026 Grenoble, France}

\author{D.~Rebreyend}
\affiliation{Universit\'e Grenoble Alpes, CNRS, Grenoble INP, LPSC-IN2P3, 38026 Grenoble, France}

\author{S.~Roccia}
\affiliation{Universit\'e Grenoble Alpes, CNRS, Grenoble INP, LPSC-IN2P3, 38026 Grenoble, France}

\author{D.~Rozpedzik}
\affiliation{Marian Smoluchowski Institute of Physics, Jagiellonian University, 30-348 Cracow, Poland}

\author{W.~Saenz-Arevalo}
\affiliation{Normandie Universit\'e, ENSICAEN, UNICAEN, CNRS/IN2P3, LPC Caen, 14000 Caen, France}

\author{P.~Schmidt-Wellenburg\orcidC{}}
\affiliation{Paul Scherrer Institut, CH-5232 Villigen PSI, Switzerland}

\author{A.~Schnabel}
\affiliation{Physikalisch-Technische Bundesanstalt, 
Abbestr. 2-12, D-10587 Berlin, Germany}

\author{E.\ P.~Segarra\orcidA{}}
\thanks{\bf The author to whom correspondence may be addressed: efrain.segarra@psi.ch, georg.bison@psi.ch}
\affiliation{Paul Scherrer Institut, CH-5232 Villigen PSI, Switzerland}

\author{N.~Severijns\orcidH{}}
\affiliation{Instituut voor Kern- en Stralingsfysica, University of Leuven, B-3001 Leuven, Belgium}

\author{K.~Svirina}
\affiliation{Universit\'e Grenoble Alpes, CNRS, Grenoble INP, LPSC-IN2P3, 38026 Grenoble, France}

\author{R.~Tavakoli Dinani}
\affiliation{Instituut voor Kern- en Stralingsfysica, University of Leuven, B-3001 Leuven, Belgium}

\author{J.~Thorne}
\affiliation{Laboratory for High Energy Physics and
Albert Einstein Center for Fundamental Physics,
University of Bern, CH-3012 Bern, Switzerland}

\author{J.~Vankeirsbilck}
\affiliation{Instituut voor Kern- en Stralingsfysica, University of Leuven, B-3001 Leuven, Belgium}

\author{J.~Voigt}
\affiliation{Physikalisch-Technische Bundesanstalt, 
Abbestr. 2-12, D-10587 Berlin, Germany}

\author{N.~Yazdandoost\orcidD{}}
\affiliation{Department of Chemistry – TRIGA site, Johannes Gutenberg University Mainz, D-55128 Mainz, Germany}

\author{J.~Zejma}
\affiliation{Marian Smoluchowski Institute of Physics, Jagiellonian University, 30-348 Cracow, Poland}

\author{N.~Ziehl\orcidI{}}
\affiliation{
Institute for Particle Physics and Astrophysics,
ETH Z\"urich,
CH-8093 Zurich, Switzerland}

\author{G.~Zsigmond}
\affiliation{Paul Scherrer Institut, CH-5232 Villigen PSI, Switzerland}

\collaboration{The nEDM collaboration at PSI}

\date{\today}

\begin{abstract}
High-precision searches for an electric dipole moment of the neutron (nEDM) require stable and uniform magnetic field environments. We present the recent achievements of degaussing and equilibrating the magnetically shielded room (MSR) for the n2EDM experiment at the Paul Scherrer Institute. We present the final degaussing configuration that will be used for n2EDM after numerous studies. The optimized procedure results in a residual magnetic field that has been reduced by a factor of two. The ultra-low field is achieved with the full magnetic-field-coil system, and a large vacuum vessel installed, both in the MSR. In the inner volume of $\sim1.4~\si{\meter}^3$, the field is now more uniform and below 300~\si{\pico\tesla}. In addition, the procedure is faster and dissipates less heat into the magnetic environment, which in turn, reduces its thermal relaxation time from $12~\si{\hour}$ down to $1.5~\si{\hour}$.
\end{abstract}

\maketitle

\section{Introduction}
\vspace{-3mm}
\label{sec:Introduction}
n2EDM is the current state of the art experiment, carrying out a high-precision search for an electric dipole moment of the neutron~\cite{Ayres2021n2EDM} at the ultra-cold neutron source~\cite{10.21468/SciPostPhysProc.5.004} of the Paul Scherrer Institute. The experiment will deliver an order of magnitude better sensitivity than previous efforts~\cite{PhysRevLett.124.081803}, down to $1\times 10^{-27} e$~\si{\centi\meter}. The experiment precisely extracts the spin precession frequency of ultra-cold neutrons in a weak magnetic field, $B_0$ and a strong electric field, $E$, via Ramsey's method of separated oscillating fields~\cite{PhysRev.78.695}. In order to reach this sensitivity, a stable and uniform magnetic field environment is critical. Thus, shielding the precession chamber from external magnetic flux is crucial.

The n2EDM precession chamber is held in a magnetically shielded room (MSR), utilizing both active and passive magnetic shielding components~\cite{10.1063/1.4894158,Ayres2021n2EDM,doi:10.1063/5.0101391,FutureAMSPaper}. The active magnetic shielding compensates external magnetic-field drifts, maintaining a constant field environment on the outside of the MSR of $\sim 1$~\si{\micro\tesla}~\cite{FutureAMSPaper}. In the MSR, passive shielding materials are used to reduce the magnetic flux to the innermost volume~\cite{doi:10.1063/5.0101391}. Passive material often used for shielding have a high magnetic permeability, such as MUMETALL$^\textrm{\textregistered}$, a soft ferromagnetic alloy of nickel and iron used in the construction of this MSR.

The MSR inner volume is almost perfectly cubic, with a side length of 2.93~\si{\meter} and internal volume of 25~\si{\meter^3} for the experimental apparatus. The outside dimensions of the MSR is $5.2~\si{\meter}\times 5.2~\si{\meter}$ horizontally and $4.8~\si{\meter}$ vertically. To achieve a quasi-static shielding factor of $\sim10^5$ at 0.01~\si{\hertz}, the MSR is composed of seven shielding layers: one aluminium layer (serving as an RF shield) and six soft magnetic layers (five layers of MUMETALL$^\textrm{\textregistered}$ and one layer, ``layer 6'', of the alloy ULTRAVAC 816$^\textrm{\textregistered}$)~\cite{doi:10.1063/5.0101391}.

The magnetic environment during data-taking for the n2EDM experiment consists of a constant, vertical magnetic field, $B_0$, of about 1~\si{\micro\tesla} in the direction of the $\pm z$-axis, produced by a cuboid coil~\cite{Ayres2021n2EDM}.

To obtain a pristine magnetic environment that is equilibrated
to the weak magnetic field $B_0\sim 1$~\si{\micro\tesla}, contributions of the residual magnetic field must be near-zero. Near-zero residual magnetic field is achieved via ``degaussing'' the passive shielding, or ``equilibrating'' it to stable field conditions, with respect to the additional $1$~\si{\micro\tesla} of the experiment.

Degaussing reduces, ideally ``erases'', the residual magnetization of a material. It is typically done by applying a strong, alternating polarity, sinusoidal magnetic field. The magnetic flux must be initially strong enough to completely saturate the material everywhere, thereby erasing the previous magnetization state. The amplitude of the oscillating field then slowly decays to zero, producing randomized magnetic domains in the material. If the field does not decay to exactly zero, i.e., if any DC current offset exists, a residual magnetization will remain. See Refs.~\cite{10.1063/1.2713433,Voigt_Jens_Measures_2013,10.1063/1.4922671,9075432} for recent reviews and efforts of degaussing MSRs. Additionally, degaussing will heat the material. During the thermal relaxation after degaussing, magnetic field drifts occur. A reduction of dissipated head would reduce the thermal relaxation time towards stable field conditions, and is therefore ideal.

To produce the alternating magnetic flux for degaussing, currents are applied to coils wound around the shielding layers. As also developed in Ref.~\cite{9075432}, n2EDM features a novel distributed coil design for a more complete degaussing of the innermost shielding layer. A successful degaussing of the MSR, with low field values of below $600~\si{\pico\tesla}$ for the entire inner volume of the MSR ($\sim 25\si{\meter}^3$), has already been demonstrated in Ref.~\cite{doi:10.1063/5.0101391}. While the residual field found there is sufficient enough to allow for a sensitive n2EDM measurement, improvements were sought due to the long duration of the degaussing procedure (3.5~\si{\hour}) and the long thermal relaxation time after a degaussing (12~\si{\hour}).

In this article, we present further improvements to the degaussing procedure, including further reducing the residual magnetic field and improving its uniformity, all while taking less time and dissipating less heat. This work presents the experimental constraints, discuss the degaussing design and procedure, and outline improvements made.
 \vspace{-2em}

\section{Experimental requirements}
\vspace{-3mm}
\label{sec:Constraints}

\begin{figure}[th!]
\centering
\includegraphics[width=0.74\columnwidth]{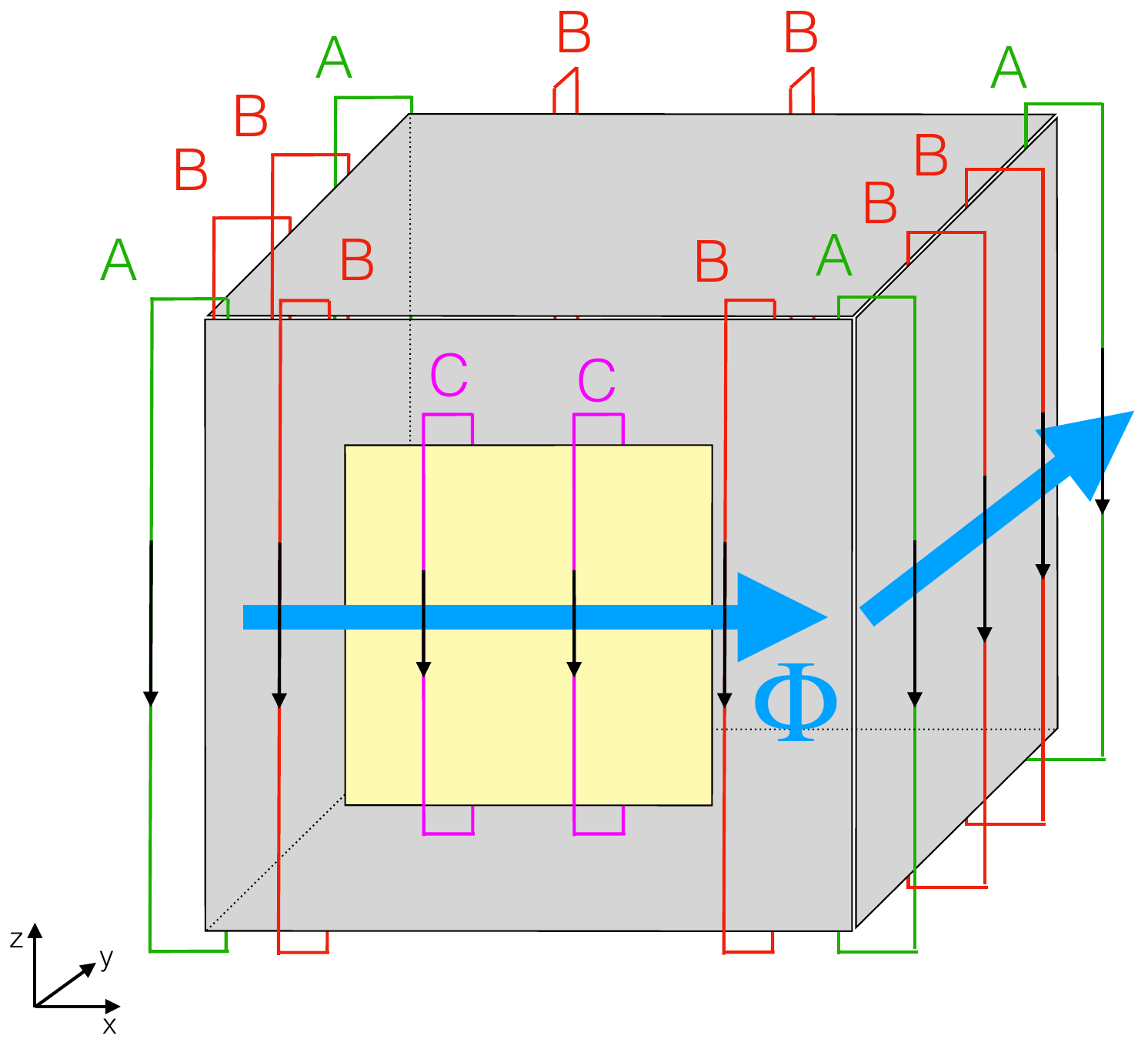}
\caption[]{
Arrangement of the degaussing coils, which produce flux around the $z$-axis
on
MSR layer 6
drawn as a cube box. The yellow square represents the access door.
Label A (green): corner coils similar on all layers;
Label B (red): additional coils only on layer 6;
Label C (purple): additional smaller coils only on the
layer 6 door.
The blue arrows indicate the direction of the
magnetic flux $\Phi$
produced inside the shielding material by a current through the indicated coils.
}
\label{fig:Degaussing-Coils}
\end{figure}

\begin{figure}[bh!]
\centering
\includegraphics[width=0.81\columnwidth]{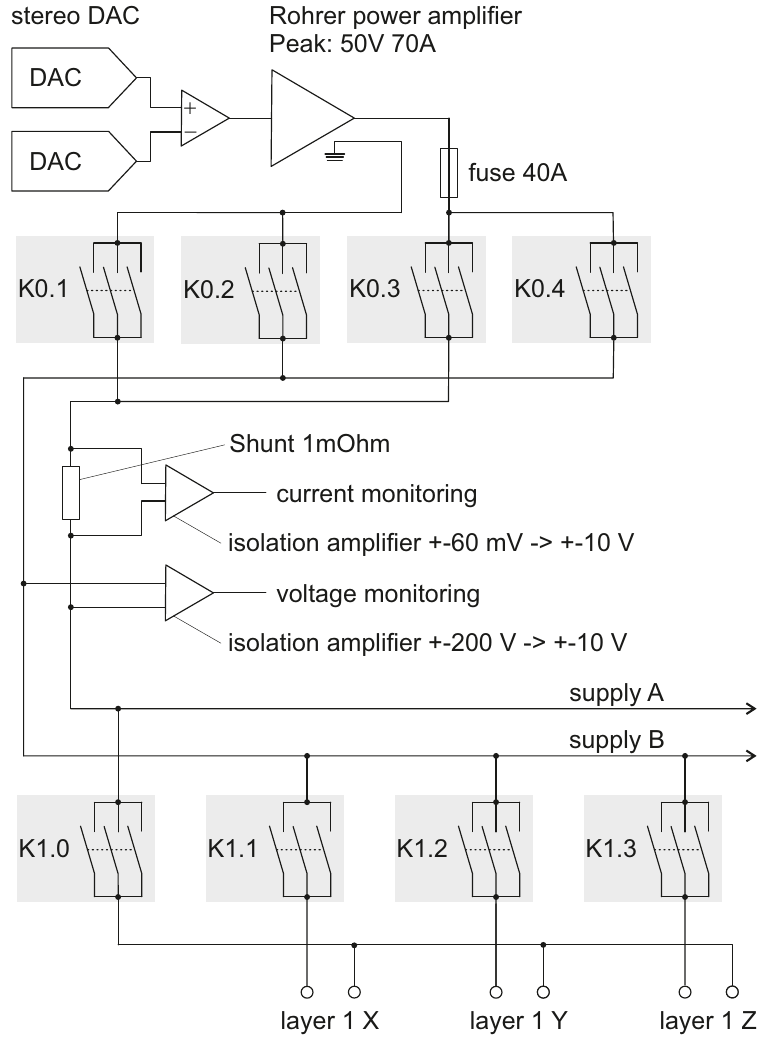}
\caption[]{
Scheme of the electrical connections from the DACs to the degaussing coils of layer 1.
The relays K0.1 to K0.4 are the main switches, which can connect the supply rails "supply A" and "supply B" in either polarity to the power amplifier.
The relays K1.1 to K1.3 are used to connect the degaussing coils of layer 1 to the supply rails, selecting which of the $x$-, $y$-, and $z$-degaussing coil is supplied with current.
In order to activate the return path for layer 1, K1.0 is closed.
In the idle state when the layer one coil is not powered, all K1 relays are open.
Identical schemes are used for layers 2 to 5.
}
\label{fig:scheme_l1}
\end{figure}

\begin{figure}[ht!]
\centering
\includegraphics[width=0.81\columnwidth]{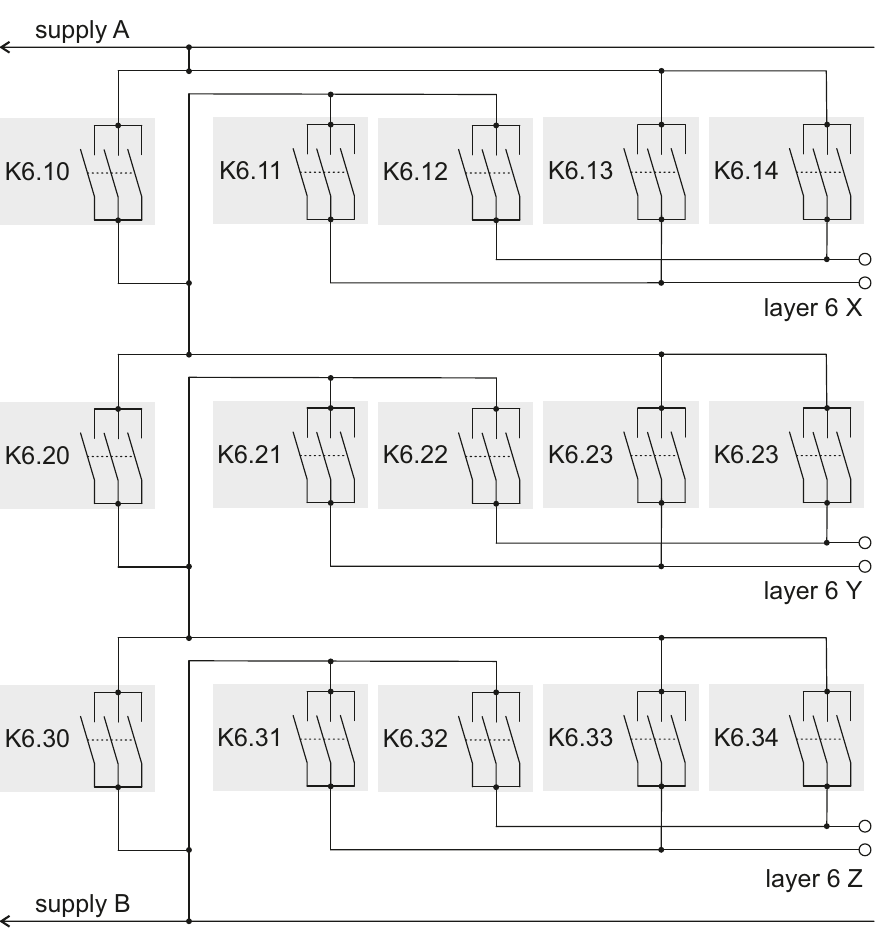}
\caption[]{
The scheme of the relay connections for layer 6 is more elaborate then for the other layers.
It enables connection to any of the three coils in series with arbitrary relative polarity.
The signal always goes from the rails supply A to supply B (see Fig.~\ref{fig:scheme_l1}).
The relays K6.10, K6.20, and K6.30 select which coils are not powered by providing a current path that bypasses the coil.
The hardware (using relay logic) does not permit all three of those relays to close simultaneously, avoiding a short circuit.
When K6.10 is open, two of the relays K6.11 to K6.14 are closed in order to select in which polarity the current runs though the x degaussing coil of layer 6.
The same scheme is used to power independently the y and z coils.
In the idle state when the layer 6 coil is not powered all K6 relays are open.
}
\label{fig:scheme_l6}
\end{figure}

In order to reach the sensitivity goal of the n2EDM experiment~\cite{Ayres2021n2EDM}, the influence of the residual field on the overall strength and non-uniformity of the $1$~\si{\micro\tesla} $B_0$ magnetic field should be minimal. The specific design requirements is that the residual field must be below 500~\si{\pico\tesla} in the central 1~$\si{\meter^3}$ with a field gradient less than 300~\si{\pico\tesla/\meter}, see Ref.~\cite{Ayres2021n2EDM}. The length of time to degauss, and the heat dissipation due to degaussing, should be minimized to maximize available measurement time under stable magnetic-field conditions.
 \vspace{-1em}

\section{Design of the degaussing system}
\vspace{-3mm}
\label{sec:Design}
In order to degauss each layer of the MSR individually, coils are installed in order to produce a magnetic flux around each spatial direction, $x,y,$ and $z$ axes. This corresponds to $x,y,$ and $z$ degaussing. In the initial setup of the n2EDM degaussing system, we followed the procedures laid out in Refs.~\cite{10.1063/1.2713433,Voigt_Jens_Measures_2013}.

The construction and installation of the coils were done by the company VAC~\footnote{VACUUMSCHMELZE GmbH \& Co. KG, Gruener Weg 37, D-63450 Hanau, Germany}, which produced the MSR. Prior to mounting, all parts of the coils of the inner MSR were checked for magnetic contamination with a 3D superconducting quantum interference device (SQUID) array installed in the BMSR-2 of Physikalisch Technische Bundesanstalt (PTB), Berlin~\cite{Bork2000}.

\begin{figure}[th!]
\centering
\includegraphics[width=0.74\columnwidth]{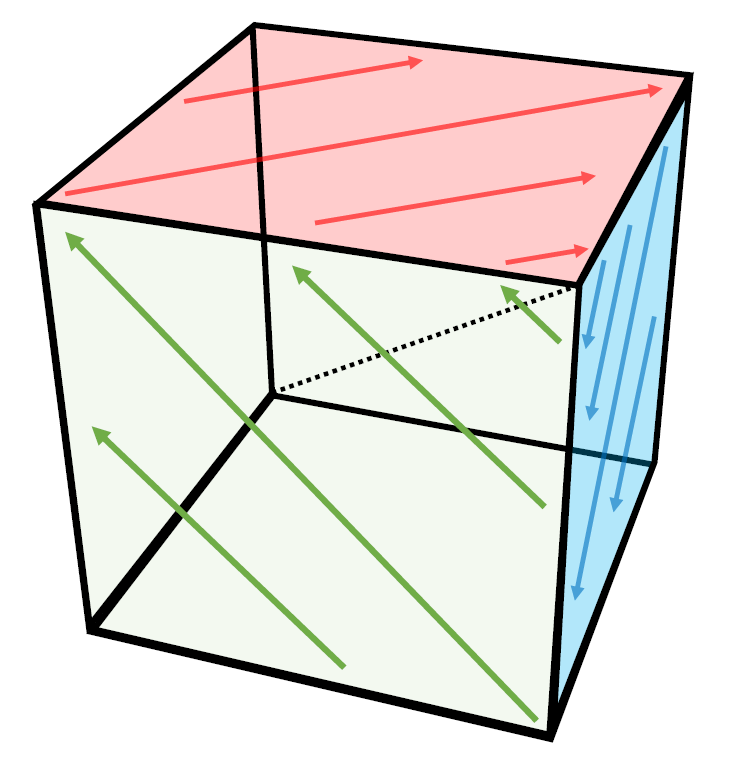}
\caption[]{
Visualization of magnetic flux across  the surface of layer 6 during a simultaneous excitation of $x,y,$ and $z$ coils. Green, blue, and red arrows indicate the flux direction on the $y$, $x$, $z$ surfaces, respectively. The dotted line indicates the corner axis around which the flux is generated.
}
\label{fig:CornerDegaussing}
\end{figure}

\vspace{-1.5em}
\subsection{Layers 1-5}
\vspace{-3mm}
\label{sec:Setup:L1-5}
The outer five MUMETALL$^\textrm{\textregistered}$ layers of the MSR (layers 1-5) follow the same coil design, with wires running along the edges, labeled A in Fig.~\ref{fig:Degaussing-Coils} for the $z-$degaussing. $x-$, $y-$, and $z-$degaussing each has four coils along the edge of each spatial direction, making twelve coils per layer in total. Each coil has seven turns fabricated out of 6~\si{\milli\meter^2} copper cables. The coils for each spatial direction are connected in series by a 6~\si{\milli\meter^2} coaxial cable, where the shielding is used as current return in order to avoid stray magnetic fields by the loop. Between RF shield of the MSR and the electrical cabinet hosting the electrical circuits, the coaxial cables are additionally shielded by nested copper tubes. They protect against interference capacitively coupling to the degaussing coils, but do not connect the ground potential of the MSR and the special EMI-shielded electrical cabinet.

Inside the cabinet, relays can connect the cables from the coils to the power amplifier. Fig.~\ref{fig:scheme_l1} shows the electrical scheme for connecting the power amplifier that is controlled by Digital-Analog-Converters (DACs) to the coils for layer 1. It allows for switching on and off the coils for the three spatial directions in a defined sequence. This is identical for layers 2-5.

\vspace{-1.5em}
\subsection{Layer 6}
\vspace{-3mm}
\label{sec:Setup:L6}
The innermost ULTRAVAC$^\textrm{\textregistered}$ layer (layer 6) utilizes a distributed coil design in each spatial direction, using more coils over the width of the walls between each corner. The additional coils along the surfaces are labeled B and C in Fig.~\ref{fig:Degaussing-Coils}, where coils C specifically cover the MSR door.

This distributed design allows a more uniform magnetic flux to be obtained, and as a consequence, less residual magnetization along the edges and in the corners. See Ref~\cite{9075432} for a detailed discussion on distributed coils.

\begin{figure}[t]
    \centering
    \includegraphics[width=0.48\textwidth]{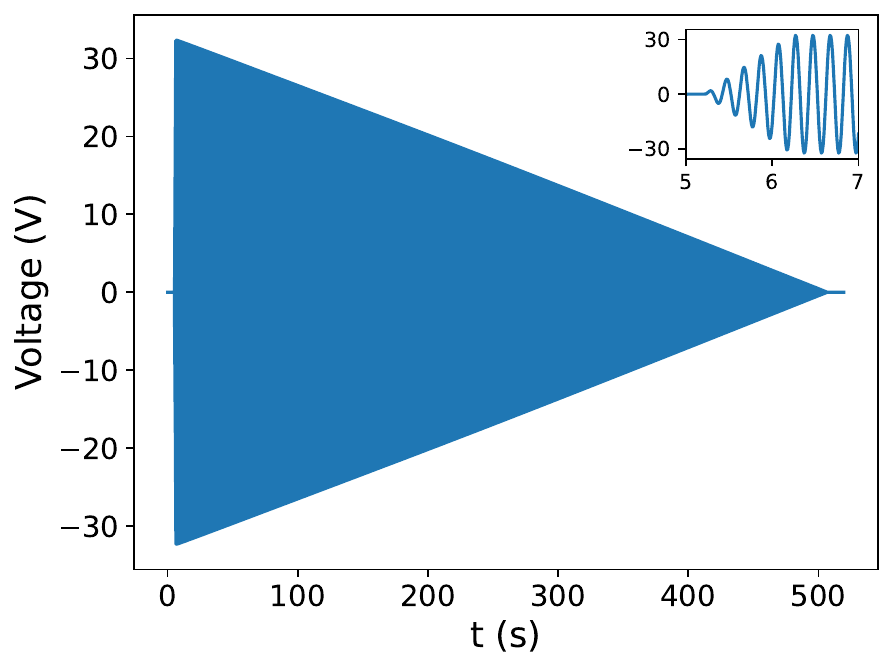}
    \caption{Voltage monitoring of the Rohrer amplifier output as a function of time with a linear increase/decrease of amplitude. The zoomed inset shows the oscillating behavior. An up-time of $1$~\si{\second}, hold-time of $1$~\si{\second}, down-time of $500$~\si{\second}, and frequency of $5$~\si{\hertz} is used.}
    \label{fig:degaussfunction}
\end{figure}

Each spatial direction, thus, has at least twelve coils -- four on each corner plus two distributed approximately equidistant along the wall. In generating the magnetic flux around the $x-$ and $z-$axis, two additional coils are used around the innermost MSR access door. Therefore, layer 6 has 40 coils in total.

In order to drive the coils in any combination of up to three spatial directions, each with independent polarities, at the same time, layer 6 also has a unique electrical scheme, see Fig.~\ref{fig:scheme_l6}. For example, a flux generated simultaneously along $x,y,$ and $z$ would correspond to a flux around one of the four corner-axes of the MSR, see Fig.~\ref{fig:CornerDegaussing}.
\vspace{-1em}

\subsection{Generating the degaussing magnetic flux}
\label{sec:Setup:Generating}

In order to degauss a ferromagnetic material with an AC magnetic field, the material must first be magnetically saturated to remove the magnetic history. After saturation, any strong AC function with alternating sign and decreasing amplitude can lead to a non-magnetic material state (excluding any DC offset of the AC field), if enough cycles are used, and if done in a zero-field environment. For the n2EDM experiment, the outer 5 layers of the MSR provide a nearly zero-field environment for layer 6 degaussing.

We control the voltage over time supplied to the degaussing coil to produce a degaussing waveform. The supplied waveform is divided into three segments: ``up-time" (time-to-peak), ``hold-time" (time-at-peak), and ``down-time" (time-to-zero). Additional parameters that can be varied are the maximal peak amplitude and the frequency. The final degaussing waveform to degauss with a finite number of cycles (i.e., finite time) is shown in Fig.~\ref{fig:degaussfunction}. Although only linear ramps have been used, one could easily introduce non-linear ramps.

To reach saturation throughout the shielding material, the frequency was reduced to 5~\si{\hertz}, which also reduces the maximal current needed to reach saturation. The disadvantage of this is a prolonged degaussing time for the same number of down cycles. And practically, a transformer cannot be used to eliminate a DC offset on the degaussing signal after the power amplifier, as the lowest frequency for commercial transformers is 7~\si{\hertz}. Instead, the DC offset of the power amplifier is measured regularly with a multimeter and is adjusted to $\sim0.01~\si{\milli\volt}$. The DC offset was originally optimized with a magnetometer installed close to the MSR-center, searching for which offset lead to the lowest residual field at this position. It was confirmed that a zero-voltage offset at the degaussing coil leads to the lowest residual field.

\begin{figure}[t]
    \centering
    \includegraphics[width=0.48\textwidth]{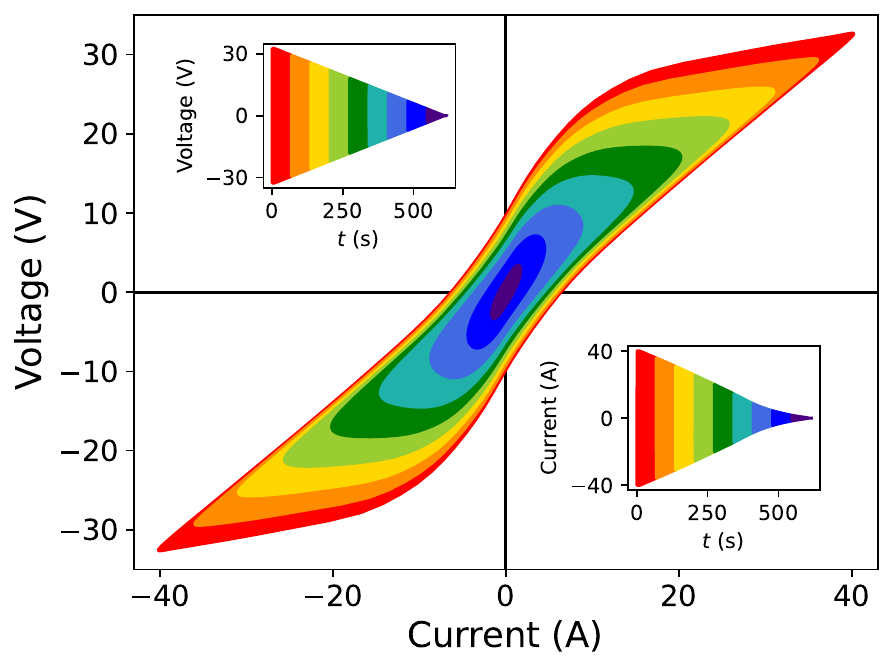}
    \caption{Voltage output of Rohrer amplifier as a function of current. The upper-left inset figure shows the voltage monitoring as a function of time, similar to Fig.~\ref{fig:degaussfunction}, and is linear as the Rohrer is operated in voltage-mode. The different colors highlight which part of the degaussing waveform populates the current-voltage space. The lower-right inset is the same but for current as a function of time. See text for details.}
    \label{fig:hysteresis_curve}
\end{figure}

The degaussing waveform is generated by the two DACs with opposite polarity and converted to a single-ended signal by a low-noise differential amplifier. A Rohrer  PA2088A 3.5~\si{\kilo\watt} power amplifier~\footnote{Rohrer GmbH,  D-81457 Munich, Germany, rohrer-muenchen.de}  is operated in voltage mode and generates a high power signal of up to $\pm \SI{50}{~V}$ proportional to the input waveform. Current and voltage monitoring is implemented using insulation amplifiers. Fig.~\ref{fig:hysteresis_curve} shows the voltage vs current monitoring while degaussing along one axis in one layer. Reaching saturation is indicated by the nearly linear relation at the tips of the voltage-current curve of Fig.~\ref{fig:hysteresis_curve}, showing the loss of the inductive part with the remaining ohmic part of the degaussing coil resistance. In practice, there will always be a small inductive part. Firstly, there are places where shielding material overlaps, and at the MSR access door, where shielding material is thicker, requiring very large currents to reach complete saturation. Additionally, for larger currents, the field lines will cross gaps between layers, and thus, use the next shielding layer as a return path. For sequential degaussing of layers 1-5, the field lines can also take the path over the two surfaces of the shielding layer that are not in the main path of the magnetic flux $\Phi$ as indicated in Fig.~\ref{fig:Degaussing-Coils}.
\vspace{-1em}

\subsection{Measuring residual magnetic fields}
\label{sec:Setup:Measuring}

In order to quantify the residual magnetic field in the MSR after degaussing, a magnetic-field mapper was used. The magnetic-field mapper can move a low-noise Bartington MAG13 three-axis
fluxgate~\footnote{Bartington Instruments Ltd, Thorney Leys Park, Witney OX28 4GE, United Kingdom, Bartington.com} along the cyclindrical coordinates $\rho,\phi,z$. The fluxgate can move between $\phi\in[-30,380]^\circ$, $\rho\in[-5,76]$~\si{\centi\meter}, and $z\in[-39.6,50]$~\si{\centi\meter}, which samples a volume larger than the one relevant  for n2EDM measurements. Due to the sampling time of the mapper in the large volume, magnetic field maps can be taken before and after degaussing for these studies, but not during.
This mapper follows a similar design as the one utilized in the nEDM experiment~\cite{PhysRevA.106.032808} but with substantial upgrades. A more detailed description of the internal coil system of n2EDM, including the magnetic field mapper used here, will be part of a forthcoming publication~\cite{FutureMapperPaper}.

We emphasize that the magnetic field mapping done in this work was accomplished with this automated mapper installed inside the MSR. This was not used in the work of Ref.~\cite{doi:10.1063/5.0101391}. Furthermore, here, the full magnetic coil system and vacuum vessel were both installed, neither of which were in place during the work of Ref.~\cite{doi:10.1063/5.0101391}.
 \vspace{-1em}

\section{Degaussing procedure}
\vspace{-3mm}
\label{sec:Degaussing_procedure}
During initial studies of the MSR, degaussing in layers 1-6 were performed identically. Working from outer layers to inner layers, each layer first had flux generated around $x$, then $y$, then $z$, independently. This is referred to as ``serial'' degaussing. Layers 1-5 only have the option to be degaussed serially due to the implemented switch setup, as shown in Fig.~\ref{fig:scheme_l1}. Layer 6 has the additional flexibility to degauss simultaneous $x,y$ and $z$, as shown in Fig.~\ref{fig:scheme_l6}. However, initially, it was degaussed serially as well. The degaussing waveform was also identical for each layer and each spatial axis: up-time of $200$~\si{\second}, hold-time of $10$~\si{\second}, down-time of $500$~\si{\second}, and frequency of $5$~\si{\hertz}. This waveform was initially chosen relying on previous works and experience. This degaussing procedure already allowed us to achieve extremely low residual field values over the inner volume of the MSR, as published in Ref.~\cite{doi:10.1063/5.0101391}.

\begin{figure}[t]
    \centering
    \includegraphics[width=\columnwidth]{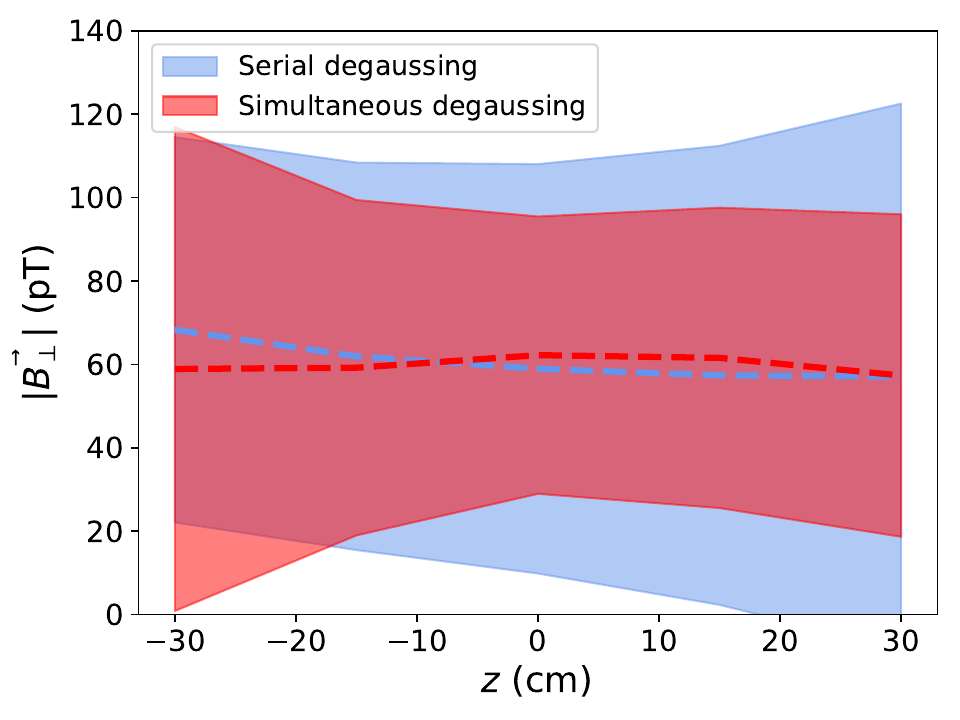}
    \caption{$\phi$-averaged transverse residual magnetic field, $B_\perp$, at $\rho=50$~\si{\centi\meter}, as a function of $z$, for the serial degaussing sequence (blue) and the simultaneous degaussing sequence with polarity $x_+y_-z_+$ (red). The dashed line shows the average field sampled in $\phi$ at $\rho=50$~\si{\centi\meter} with fixed $z$ points, and the envelope is $2\sigma$ spread around the average at each $z$.}
    \label{fig:serial_vs_simultaneous}
\end{figure}

Yet, with the aforementioned degaussing waveform and serially degaussing of each layer, a full degaussing of the MSR takes roughly 3.5~\si{\hour}. This procedure (which in the rest of this paper is referred to as 
`previous sequence') also introduces significant heat into the MSR. After a full degaussing, the residual magnetic field took many hours to thermally equilibrate (see discussion later). 

With this in mind, a series of studies were performed to optimize the degaussing procedure with two goals: (1) reduce the time spent and heat dissipated of degaussing, and (2), minimize, and make more uniform, the residual magnetic field. In order to test the effectiveness of different degaussing procedures independently, layer 6 is magnetized before each degaussing. This is achieved by turning on the $B_0$ field to 10~\si{\micro\tesla}, magnetizing significantly in $\phi$ and $z$ (mapper coordinates). This is a factor of 10 higher than the nominal $B_0$ field setting for n2EDM operation~\cite{Ayres2021n2EDM}. To magnetize in $\rho$, a large DC current is sent through the layer 6 $y$-axis degaussing coils. After each degaussing test, the MSR was able to thermally equilibrate over a few hours, depending on the test, see below.

\begin{figure}[t]
    \centering
    \includegraphics[width=\columnwidth]{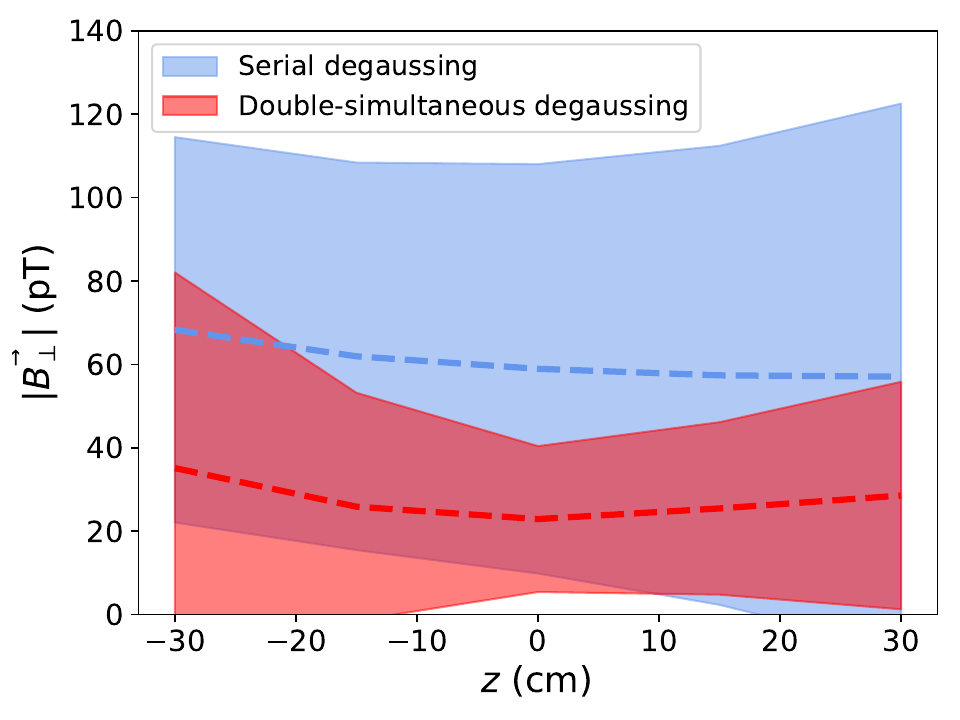}
    \caption{Same as Fig.~\ref{fig:serial_vs_simultaneous} but for the serial degaussing sequence (blue) and the double-simultaneous degaussing sequence with polarity $x_+y_-z_-$ then $x_+y_-z_+$ (red).}
    \label{fig:serial_vs_doublesimultaneous}
\end{figure}

\begin{figure*}[ht!]
    \centering
    \includegraphics[width=\columnwidth]{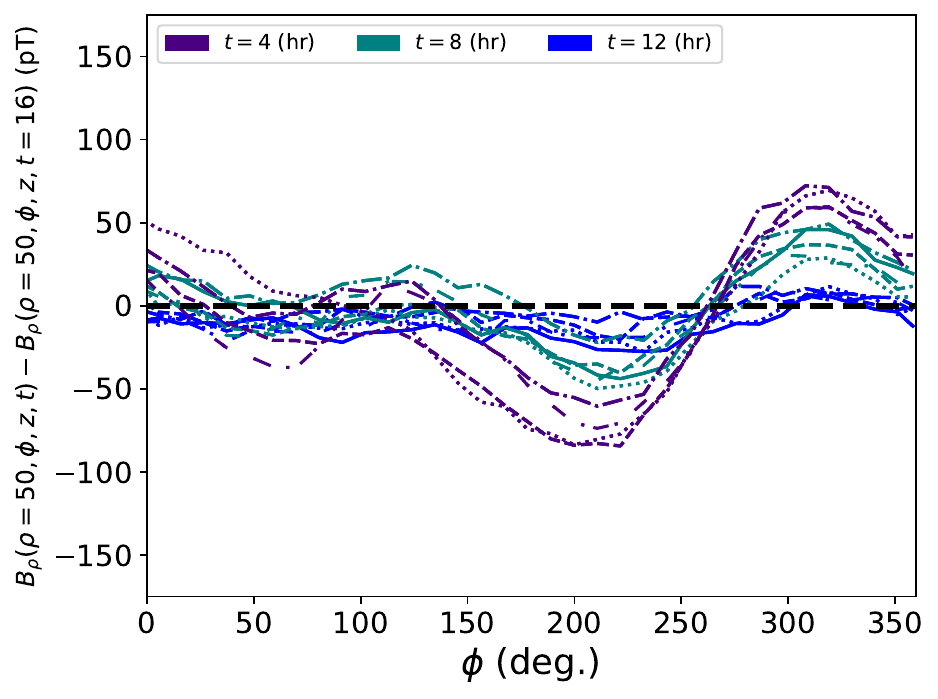}
    \includegraphics[width=\columnwidth]{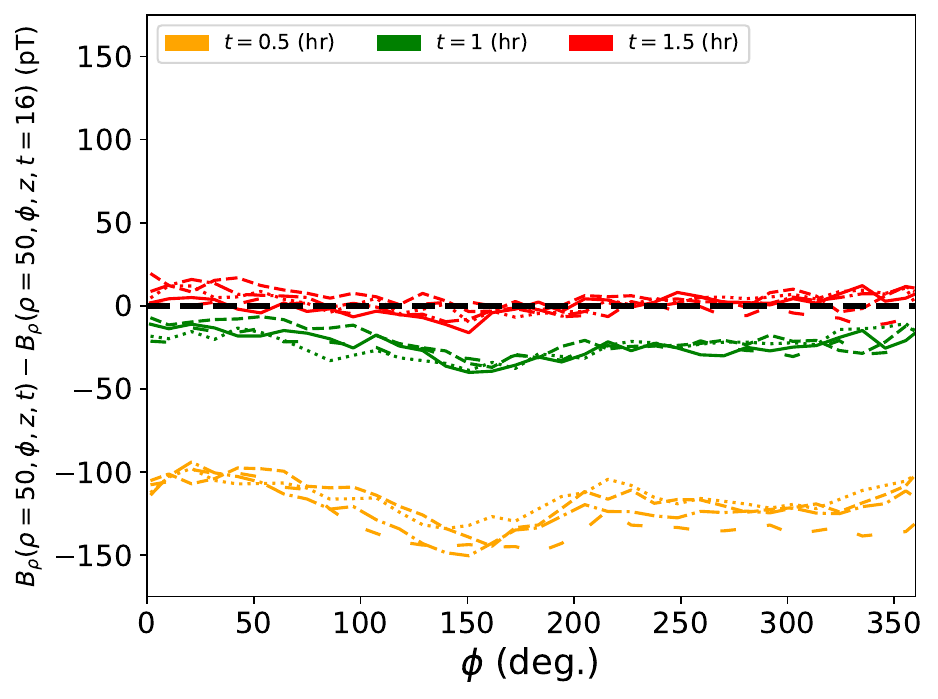}
    \caption{Thermal relaxation effect on the residual magnetic field $B_\rho$, after a full degaussing. (Left): Using the previous sequence, the difference of the magnetic field at various wait times [4 (purple), 8 (teal), and 12 (blue) hours] and the field after 16 hours, at $\rho=50$~\si{\centi\meter} as a function of $\phi$ for different $z$ positions [(solid): $z=-30$~\si{\centi\meter}, (dotted): $z=-15$~\si{\centi\meter}, (dashed): $z=0$~\si{\centi\meter}, (dotted-dashed): $z=15$~\si{\centi\meter}, (loosely-dashed): $z=30$~\si{\centi\meter}].  The different line styles indicate the various $z$ positions sampled. (Right): The same but for the optimized degaussing sequence for wait times [0.5 (yellow), 1.0 (green), and 1.5 (red) hours]. Already after 1.5~\si{\hour}, the field has reached equilibrium with the state after 2~\si{\hour}, and, even initially, the field is much more uniform in $\phi$. Uncertainty has been omitted for clarity, but the spread in the different lines of the same color give a sense of the field non-uniformity in the precession volume.}
    \label{fig:thermal_relaxation}
\end{figure*}
\begin{figure}[t]
    \centering
    \includegraphics[width=\columnwidth]{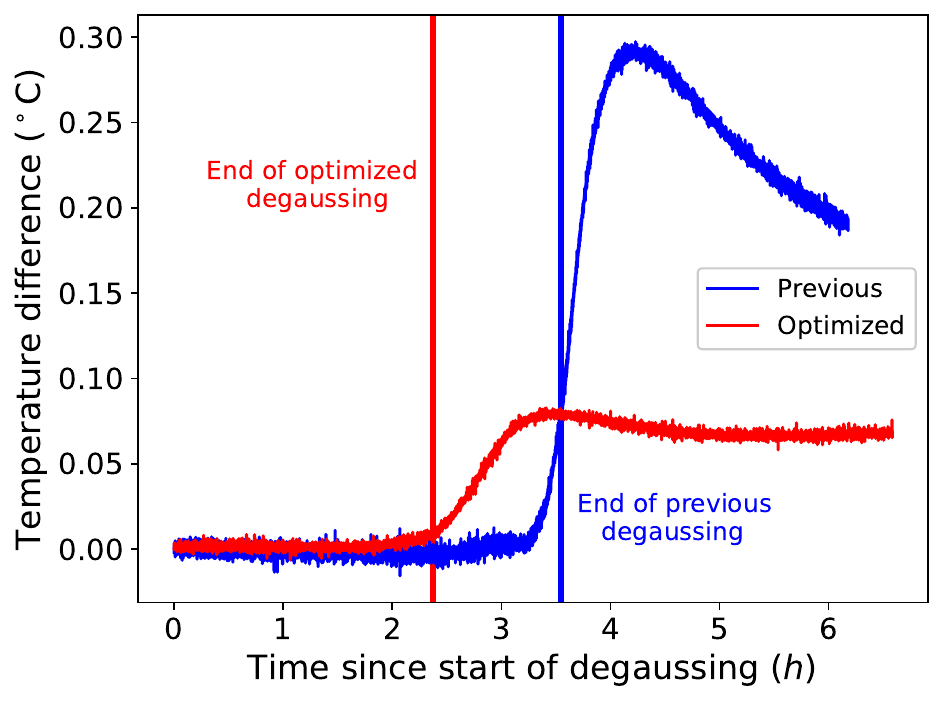}
    \caption{Temperature drift after starting a degaussing with the optimized sequence (red) and previous sequence (blue), measured by a thermocouple mounted on the inside of the MSR. The solid lines indicate when the degaussing sequence is finished (red at $\sim2.5~\si{\hour}$ and blue at $\sim3.5~\si{\hour}$). With the previous sequence, there is a significantly longer decay time until thermal stability ($~\sim12~\si{\hour}$), whereas the new sequence reaches stability in about $1.5~\si{\hour}$ after the degaussing finishes, as supported by Fig.~\ref{fig:thermal_relaxation}.}
    \label{fig:temp_drift}
\end{figure}

\subsection{Optimizing degaussing waveform}
\label{sec:Optimization_l1-5}
Since layers 1-5 only have the option to be degaussed serially, the obvious candidate for improvement of time-spent and heat-output is varying the degaussing waveform. The time and heat can be reduced with a shorter or smaller-amplitude degaussing waveform. Numerous studies were done varying the amplitude, up-time, down-time, and hold-time. For each test, layers 1-6 were always degaussed in the same manner.

It was found that an up-time of 1~\si{\second}, a hold-time of 1~\si{\second}, and a down-time of 500~\si{\second} (see Fig.~\ref{fig:degaussfunction}) produced similar residual magnetic field strengths as the previous sequence. This suggests that magnetic saturation can be reached quickly (in 1~\si{\second}) and reliably with the setup of our system. This reduces the overall degaussing time from roughly 3.5~\si{\hour} to 2.5~\si{\hour}. In addition, these changes significantly reduce the heat output, as the coils have the maximum current for only a tenth of the time as previous. Decreasing the amplitude worsened the residual magnetic field, and thus, was not changed.

\subsection{Optimizing degaussing order}
\label{sec:Optimization_l6}
Layer 6 is the innermost layer, and therefore, has the largest effect on the residual field. Initially, the order of the degaussing axes was varied to investigate the impact on the strength of the residual magnetic field. Instead of degaussing layer 6 in order of $x,y,z$, a degaussing order of $y,x,z$ was studied. This allows for a double-degaussing around the MSR access door in the last two steps, where it is assumed to have worse field uniformity due to the discontinuities in the MUMETALL$^\textrm{\textregistered}$. However, no significant improvement of the residual magnetic field with a $y,x,z$ degaussing was found, while not ending the sequence along $z$ always led to a worse residual field. 

Layer 6 is also equipped with the possibility to degauss two or three axes simultaneously. This is the most appealing candidate for further reducing the residual magnetic field. Simultaneously degaussing around $x,y,$ and $z$ (i.e., producing magnetic flux around a corner-axis of the MSR) has a large potential to reduce the necessary time and heat impact. As layer 6 also has the flexibility to independently vary the polarity of the degaussing in each axis, there are 8 different flux variations: (1) $x_+y_+z_+$, (2) $x_+y_+z_-$, (3) $x_+y_-z_+$, (4) $x_+y_-z_-$, (5) $x_-y_+z_+$, (6) $x_-y_+z_-$, (7) $x_-y_-z_+$, (8) $x_-y_-z_-$. The last four combinations are just an inversion of the current direction of the first four. For an AC degaussing, the overall sign should not have an impact if the down-time is long enough, and if the DC offset is close enough to zero.

Using the degaussing waveform with the smaller up- and hold-time as mentioned in Section~\ref{sec:Optimization_l1-5}, each of the 8 possible degaussing combinations were tested. In between each test, layer 6 was re-magnetized and the MSR was able to thermally equilibrate over 1.5~\si{\hour}. The shorter time necessary to equilibrate already reflects the reduced heat output of simultaneous degaussing.

It was found that degaussing $x,y,$ and $z$ simultaneously produced a similar residual magnetic field as the serial procedure, but with better uniformity. This is shown in Fig.~\ref{fig:serial_vs_simultaneous} by the reduced envelope of $\small{|\vec{B}_\perp|=\sqrt{B_x^2 + B_y^2}}$, but similar central value. We compare the residual field at $\rho=50~\si{\centi\meter}$, as this spans the inner-volume relevant for n2EDM. $|\vec{B}_\textrm{tot}|$ is not plotted as the fluxgate DC-offset for $B_z$ was not determined here (see Fig.~\ref{fig:comparing_old_new} for a high-resolution map that included the offset correction). It also takes significantly less time and dissipates less heat. However, this improved result was only achieved for the polarity configuration $x_+y_-z_+$. The other polarity configurations yielded larger residual magnetic fields, but all had a significantly lower heat output than the previous sequence.

In order to try to reduce the residual magnetic field even more, two subsequent simultaneous degaussings were studied -- i.e., first $x_+y_+z_+$ then $x_+y_-z_+$. A total of 12 tests were performed, using all combinations of two subsequent simultaneous degaussings.

Iterations that ended with the configuration $x_+y_-z_+$ resulted in the smallest residual magnetic field. It was found that the sequence $x_+y_-z_-$ then $x_+y_-z_+$ performed the best, as shown in Fig.~\ref{fig:serial_vs_doublesimultaneous}. This is likely due to generating magnetic flux in perpendicular directions through sides of the MUMETALL$^\textrm{\textregistered}$ and ULTRAVAC$^\textrm{\textregistered}$ that have more imperfections, most significantly at the sides of the MSR access door.

Double-degaussing using the same sequence (i.e., the same polarity configuration) did not yield improvements. It was also observed that additional simultaneous degaussings (i.e., more than a sequence of 2) did not reduce the residual magnetic field further. Thus, we found no need to do further degaussings of layer 6 beyond the two.

With the double simultaneous degaussing ($x_+y_-z_-$ then $x_+y_-z_+$) and the shorter degaussing waveform, the time spent to degauss only layer 6 was reduced from 35.5~\si{\minute} to 16.7~\si{\minute}, with a smaller residual magnetic field, and generated less heat as compared to the previous degaussing waveform. This marks a simultaneous improvement of all optimization criteria in the degaussing procedure.

\subsection{New degaussing procedure}
\label{sec:New_procedure}
The optimized degaussing procedure for the full MSR is now: degauss layers 1-5 serially (in $x,y,$ then $z$) and degauss layer 6 with a double simultaneous sequence of $x_+y_-z_-$ then $x_+y_-z_+$. All layers utilize the shorter degaussing waveform with an up-time of 1~\si{\second}, a hold-time of 1~\si{\second}, a down-time of 500~\si{\second}, and a frequency of 5~\si{\hertz}. The total time to fully degauss all layers of the MSR with the new procedure is roughly 2.4~\si{\hour}, down from roughly 3.5~\si{\hour}, excluding time for thermal relaxation.

In Fig.~\ref{fig:thermal_relaxation}, the relaxation time of the residual magnetic field with the optimized procedure is compared to the previous degaussing procedure. With the less time taken to degauss and the less heat output, the optimized procedure yields a magnetic configuration that is already thermally stable after 1.5~\si{\hour}. This is an order of magnitude faster than the time to achieve thermal relaxation for the previous degaussing procedure (previously $~\sim12~\si{\hour}$). Similarly, Fig.~\ref{fig:temp_drift} shows the temperature difference of a sensor in the MSR after a degaussing has started. With the new sequence, there is a factor of $\sim4$ less rise in temperature, and a quicker time to thermal stability.

\begin{figure}[t]
    \centering
    \includegraphics[width=\columnwidth]{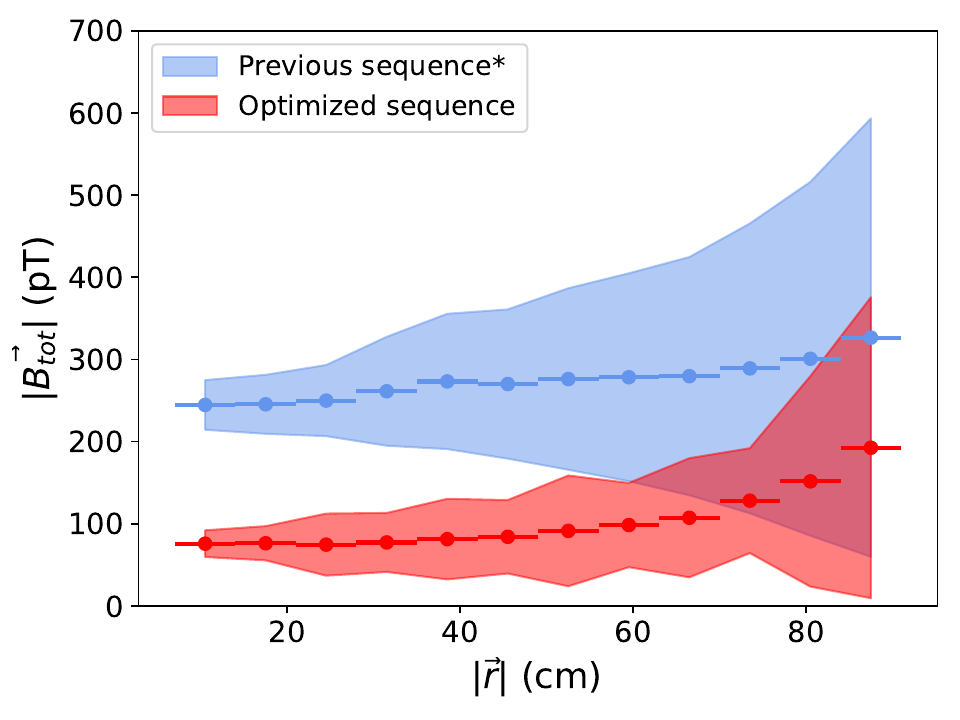}
    \caption{Average total residual magnetic field $|\vec{B}_\textrm{tot}(\vec{r})|=\sqrt{B_x^2+B_y^2+B_z^2}$, as a function of distance from center of the vacuum tank, for the serial degaussing sequence (blue) and the optimized sequence (red). The envelope is $2\sigma$ of the average field sampled in $|\vec{r}|=7~\si{\centi\meter}$ bins, where the bins are indicated by the horizontal bars. The optimized sequence includes a fluxgate correction for $B_z$ ($\sim$90~\si{\pico\tesla}), where the correction was extracted by rotating the fluxgate 180$^\circ$. The previous sequence (*) did not measure this offset, due to a mechanical issue during the measurement, however the same correction that was extracted in the optimized sequence was applied here as well. This leaves an undetermined offset, added in quadrature, still possible to the blue curve.}
    \label{fig:comparing_old_new}
\end{figure}

As compared to the previous serial degaussing of layers 1-6 with the longer degaussing waveform, the new procedure produces a substantially smaller residual field that is significantly more uniform in the innermost volume. For $|\vec{r}|$ up to $50~\si{\centi\meter}$, the new procedure has at least a factor of 2 smaller residual field, and is a factor of 2 more uniform (less spread). Fig.~\ref{fig:comparing_old_new} illustrates the improvement of the average residual magnetic field for an inner volume of $\sim1.4~\si{\meter}^3$ for the simultaneous degaussing achieved as a result of this work, compared to the serial degaussing of layer 6 used previously. 

 \vspace{-1em}
\section{Conclusions}
\vspace{-3mm}
\label{sec:Conclusions}
We developed an optimized degaussing procedure of the magnetically shielded room for n2EDM, the next-generation experimental search of the neutron electric dipole moment at the Paul Scherrer Institute~\cite{Ayres2021n2EDM}.

This procedure utilizes distributed coils, whose novel electrical design allows for producing a magnetic flux simultaneously across $x,y,$ and $z$ axes in the innermost, passive-shielding layer.

This resulted in a residual magnetic field was reduced down to below 300~\si{\pico\tesla} in the inner spin precession volume ($\sim1.4~\si{\meter}^3$) of the n2EDM experiment, despite the presence of the large experimental components inside the MSR. Even more, the residual field has been made more uniform, and all while taking less time to degauss and inputting less heat into the magnetic environment, allowing for faster thermal relaxation.

While a successful degaussing of the MSR was already demonstrated in Ref.~\cite{doi:10.1063/5.0101391}, with low residual field values (below $100~\si{\pico\tesla}$ in the inner $1~\si{\meter}^3$), the field values found there and here are not directly comparable. For one, the magnetic mapper described here was not used in Ref~\cite{doi:10.1063/5.0101391}. Additionally, the inner MSR now houses many components for the experiment, most significantly, a large vacuum tank, which contributes to the overall residual magnetic field. What is comparable are the results of Fig.~\ref{fig:comparing_old_new}, which showcase the performance of the previous and optimized degaussing sequence, with the same experimental components installed and same measurement procedure.

Looking to the future, further studies are planned to investigate the limits on how-small and how-uniform a residual magnetic field can be in such a large volume. These studies will include testing variable current amplitude or frequency during degaussing, such as Ref.~\cite{HeidelbergMSR} performed.
\vspace{-1em}

\section{Acknowledgements}
\vspace{-5mm}
Excellent technical support by Michael Meier and Luke Noorda is acknowledged. We also especially acknowledge the excellent  construction work of the group for magnetically shielded rooms of the company VAC - Vacuumschmelze, Hanau; namely of Lela Bauer, Markus Hein, Maximilian Staab, and Michael Wüst. Various PSI LOG groups supported the electrical components construction. The magnetic field mapper has been designed and installed by the LPSC mechanical department; namely Johann Menu.

Support by the Swiss National Science Foundation Projects 200020-188700 (PSI), 200020-163413 (PSI), 200011-178951 (PSI), 200021-204118 (PSI), 172626 (PSI), 169596 (PSI), 200021-181996 (Bern), 200441 (ETH), and FLARE 20FL21-186179, and 20FL20-201473 is gratefully acknowledged.
This project has received funding from the European Union’s Horizon 2020 research and innovation programme under the Marie Skłodowska-Curie grant agreement No 884104.
This work is support by the DFG (DE) by the funding of the PTB core facility center of ultra-low magnetic field KO 5321/3-1 and TR 408/11-1.
The LPC Caen and the LPSC Grenoble acknowledge the support of the French Agence Nationale de la Recherche (ANR) under reference ANR-14-CE33-0007 and the ERC project 716651-NEDM.
University of Bern acknowledges the support via the European Research Council under the ERC Grant Agreement No. 715031-Beam-EDM.
The Polish collaborators wish to acknowledge support from the National Science Center, Poland, under grant No. 2018/30/M/ST2/00319, and No. 2020/37/B/ST2/02349, as well as by the Minister of Education and Science under the agreement No. 2022/WK/07.
Support by the Cluster of Excellence ‘Precision Physics, Fundamental Interactions, and Structure of Matter’ (PRISMA+ EXC 2118/1) funded by the German Research Foundation (DFG) within the German Excellence Strategy (Project ID 39083149) is acknowledged.
Collaborators at the University of Sussex wish to acknowledge support from the School of Mathematical and Physical Sciences, as well as from the STFC under grant ST/S000798/1.
This work was partly supported by the Fund for Scientific Research Flanders (FWO), and Project GOA/2010/10 of the KU Leuven.
Researchers from the University of Belgrade acknowledge institutional funding provided by the Institute of Physics Belgrade through a grant by the Ministry of Education, Science and Technological Development of the Republic of Serbia.
\vspace{-1em}

\section{References}
\vspace{-5mm}
\label{sec:References}
\bibliography{references/degauss-references}

\end{document}